\documentclass[6pt, preprint2]{aastex}
\usepackage[]{natbib}

\begin{document}

\title{Distances of Quasars and Quasar-Like Galaxies: Further Evidence that QSOs may be Ejected from Active Galaxies}

\author{M.B. Bell\altaffilmark{1}}
\altaffiltext{1}{Herzberg Institute of Astrophysics,
National Research Council of Canada, 100 Sussex Drive, Ottawa,
ON, Canada K1A 0R6;
morley.bell@nrc.ca}

\begin{abstract}

If high-redshift QSOs are ejected from the nuclei of low-redshift galaxies, as some have claimed, a large portion of their redshift must be intrinsic (non-Doppler). If these intrinsic components have preferred values, redshifts will tend to cluster around these preferred values and produce peaks in the redshift distribution. Doppler ejection and Hubble flow components will broaden each peak. Because ejection velocities are randomly directed and Hubble flow components are always positive, in this model all peaks are expected to show an asymmetry, extending further out in the red wing. If peaks are present showing this predicted asymmetry, it can lead directly to an estimate of quasar distances. Using two quasar samples, one with high redshifts and one with low, it is shown here that not only do all peaks in these two redshift distributions occur at previously predicted preferred values, they also all show the predicted extra extension in the red wing. For the low and high redshift samples the mean cosmological components are found to be z$_{c} \sim 0.024$ and $\sim 0.066$, respectively. The difference can be explained by the improved detection limit of the high redshift sample. These results offer further evidence in favor of the model proposing that QSOs are ejected from active galaxies. 

\end{abstract}

\keywords{galaxies: active - galaxies: distances and redshifts - galaxies: quasars: general}

\section{Introduction.}

There has been a considerable amount of evidence presented in recent years suggesting that many QSOs are ejected from the nuclei of active galaxies \citep{arp98,arp99,bel02a,bel02b,bel02c,bel04,bur95,bur96,bur97a,bur97b,bur99,bur04,chu98,lop02,lop04}. In this model most have redshifts that are much larger than the parent object and this excess is usually assumed to represent an intrinsic component. At least two intrinsic redshift models have been presented which predict exact values for the preferred redshifts. However, one of these, the so-called Karlsson log(1+z) model \citep{bur01}, has failed to be confirmed in several recent attempts \citep{haw02,bel03a,bel04}. Therefore, preferred redshifts in quasars are assumed here to be defined by equation 1, hereafter referred to as the quasar intrinsic redshift equation and given by the relation

z$_{iQ}$ = z$_{f}$[$N - M_{N}$] ----------------------(1)

where z$_{f} = 0.62\pm0.01$ is the intrinsic redshift constant, $N$ is an integer and $M_{N}$ varies with $N$ and is a function of a second quantum number $n$. For $N >4$, $M_{N}$ can only have the two values, 0.0 and 0.062, if it is assumed that intrinsic $blueshifts$ are not allowed. The value of the intrinsic redshift constant, z$_{f}$, was determined in several previous studies \citep{bel02c,bel03b,bel03c}. The above equation is defined more completely by eqn A1 of \citet{bel03b}.

For $N$ = 2, $M_{2}$ was determined empirically using the positions and redshifts of the 15 QSOs \citep{bur99} located within 50\arcmin  of the Seyfert galaxy NGC 1068 \citep{bel02a,bel02b,bel02c}. It was found that 12 of these QSOs appear to have been ejected from the nucleus of NGC 1068 in four, separate, similarly structured triplets, whose ejection directions rotated with the same angular velocity as the nucleus of the galaxy. It was then realized that earlier results \citep{bur68,bur90} could also be fitted to eqn 1 if $N$ = 1. The $M_{N}$ relations for higher $N$-values were then obtained simply by extrapolating from the $M_{1}$ and $M_{2}$ relations, as shown in Table 1 (see also \citet{bel03b}). It was also found \citep{bel02c,bel03a} that the intrinsic redshift values predicted by these relations agreed with the peaks in the redshift distribution of the 574 quasars studied by \citet{kar71,kar77}. Since these sources represented independent data, this was taken as a confirmation of the NGC 1068 results.

\begin{deluxetable}{ccccc}
\tabletypesize{\scriptsize}
\tablecaption{Parameters associated with different values of $N$ in Eqn. 1 \label{tbl-1}}
\tablewidth{0pt}
\tablehead{
\colhead{$N$} & \colhead{$M_{N}$} &  \colhead{n} & \colhead{Data Source} & \colhead{Reference} 
}
\startdata
1 &  n  & 0,1,2,3...9  & quasars (z $< 0.6$) & \citet{bur90} \\
2 & n(n+1)/2  &  0,1,2,3,4,5. & QSOs near NGC 1068 & \citet{bel02c} \\
3 & [p(p+1)/2]\tablenotemark{a} & 0,1,2,3. & extrapolated from $N$=1 and 2. & 
\citet{bel02c} \\
4 & [q(q+1)/2]\tablenotemark{b} & 0,1,2. & extrapolated from $N$=1, 2 and 3. & 
\citet{bel02d} \\
\enddata 

\tablenotetext{a}{p = n(n+1)/2}
\tablenotetext{b}{q = p(p+1)/2}

\end{deluxetable}

 It has also been demonstrated \citep{tif96,tif97,bel02d,bel03a,bel03b,bel03} that galaxy redshifts may also contain an intrinsic component. Although in most cases these are quite small (z$_{i} < 0.001$), intrinsic components in quasar-like galaxies (N-type, AGN, BL Lac-type, etc.) can be as least as high as z = 0.062. Equation B1 of \citet{bel03b} is assumed here to define the preferred redshift components in galaxies. It predicts the discrete velocities reported by Tifft, and remarkably, its higher values fit exactly onto the lowest quasar intrinsic redshift in each corresponding $N$ group \citep{bel02d}.

\section{The Decreasing Intrinsic Redshift Model}

Since the term \em local model \em can be misleading, by implying that all quasars are local, the term \em decreasing intrinsic redshift model \em (DIR model) will hereafter be used. Several of the characteristics of the DIR universe also need to be pointed out. Those who believe in the standard, or cosmological redshift, model may not realize that the DIR universe is in many ways quite similar. Since the intrinsic redshifts have been found to be superimposed on top of the Hubble flow \citep{bel03c}, there appears to be no need to abandon the Big Bang in this model. However, inflation may be in trouble if it suggests that $all$ the density structure in the Universe (galaxies, clusters, etc.) was preset during the inflationary period. In the DIR model, galaxies are produced continuously throughout the entire age of the Universe and the QSO stage represents just the first, very short-lived, stage ($\sim10^{7}$ - $10^{8}$ yrs) in the evolution of a galaxy. In this model QSOs are assumed to be born out of active galaxies and are the seeds from which new galaxies form. As such they are then just \em baby \em galaxies. It is also assumed that they begin to accumulate an associated nebulosity, or \em host, \em immediately after ejection. The QSO distribution in the DIR universe is expected to be as homogeneous as is the distribution of normal galaxies. QSOs differ mainly in that they are several magnitudes less luminous (the conclusions that quasars are super luminous, and reach a peak density near z = 2, come entirely from the assumption that their redshifts are cosmological). QSO intrinsic faintness means that those detected to date will be located relatively nearby because more distant ones will have apparent magnitudes that fall below today's detection limits. The only distant quasars detected to date will be those that are lensed by an intervening galaxy or cluster. The main difference between the two models thus appears to be that new galaxies are born throughout the entire life of the DIR universe, while in the conventional model they are all assumed to be formed very early. That young galaxies are initially sub-luminous (relative to mature galaxies) in the DIR model is not significantly different from the standard model. Although in the DIR model galaxies are also born with a significant intrinsic component in their redshifts (the QSO stage), most of this disappears quickly as they evolve. Otherwise, there may be no obvious reason to indicate that galaxy evolution in the DIR model differs significantly from that in the standard model.

\section{Results}

Because, in the DIR model, QSOs are too faint to be detected at large distances with current sensitivities, their cosmological redshift components should be quite small relative to their intrinsic redshifts. In the DIR model the redshift distribution is therefore essentially an intrinsic redshift one. For large quasar source samples like the Sloan Digital Sky Survey (SDSS), there is no practical way to remove either the Doppler redshift components produced by the Hubble flow, or the randomly directed ejection velocities, from the measured redshifts. Thus the shape of the individual peaks in the redshift distribution will contain information on both the distances of the sources and the size of the ejection velocities. All peaks are then expected, (1) to be \em symmetrically \em broadened by randomly directed (positive and negative) ejection velocity components, and, (2) to be \em asymmetrically \em broadened by Hubble flow components. The latter will only extend the high redshift side of each peak because the Hubble flow components can only be positive.


Here the redshift peaks in two separate redshift samples are examined. The first is a high-redshift sample taken from SDSS data and the second is a low-redshift sample from \citet{hew93}. The detection limit is significanly better in the SDSS sample and, as a result, it is expected to contain more distant objects.

\subsection{The High-Redshift Sample}

In Fig 1 the distribution of the approximately 5000 quasar emission line redshifts from the SDSS between z = 2.4 and 4.8 \citep{ogu03,bel04} has been plotted as a bold histogram. Several peaks are clearly present in the distribution. The vertical dashed lines represent the mean of the two closely spaced intrinsic redshifts in each $N$-group predicted by eqn 1 (eqn A1 of \citet{bel03b}) which have values of 2.418 and 2.480 for $N$ = 4, 3.038 and 3.10 for $N$ = 5, 3.658 and 3.72 for $N$ =6, and 4.278 and 4.34 for $N$ = 7. The predicted redshifts are also listed in Table 1 of \citet{bel04}. Although selection effects have been suggested to explain the valleys at z = 2.7 and 3.5 \citep{ric02}, this has previously been discussed in some detail \citep{bel04} and it is assumed here that the selection effects have essentially been corrected for by adjustments to the target selection algorithm which were carried out before most of these data were obtained. Other peaks of similar strength that fall below z = 2.4, and that are coincident with predicted preferred lines at z = 0.62 and 1.24, are unexplained by selection effects \citep{bel04}, as is the valley near z = 4.1. Furthermore, the peak structure (which is periodic in z = 0.62) may continue to even higher redshifts. The distribution of high-redshift objects with x-ray detection from \citet{bra03} shows peaks at z = 4.34 and 4.96 which represent the $N$ = 7 and $N$ = 8 harmonics of 0.62, respectively.
 
 However, one might also ask whether the peaks and dips in the distribution might simply be due to random fluctuations. If quasars are uniformly distributed in space and their sampling is reasonably complete, there is no reason to suspect that fluctuations in their number distribution with redshift will be bigger than random sampling will produce. For the 5000 sources in Fig 1 these fluctuations must be much smaller than the peaks and valleys visible in the plot. The 1 sigma errors expected for this sample size have been included in Fig 1. It is therefore concluded that the peaks and valleys in the distribution are not due to statistical fluctuations, but are due to either a selection effect, or to some physical phenomenon, such as preferred intrinsic values in the DIR model as suggested here, or a periodic source density clumping in space in the cosmological redshift model. The fact that they occur at those redshifts that were predicted by a \em previously determined \em intrinsic redshift model seems to suggest that that might be the most likely explanation. That, and the fact that a \em periodic \em density clumping with cosmological z in the Universe is difficult to explain. For the source distribution in Fig 1, the decrease in number with increasing redshift is easily explained in both models by the detection limit. Perhaps most significant of all is the fact that the predicted preferred redshift values were obtained from completely independent data and were found and made available to the community long before the SDSS data in Fig 1 were published.

In Fig 1, the four dashed curves, all with similar profiles, are attempts to show, qualitatively, the shape of each individual peak. Their sums combine to reproduce the histogram as closely as possible, and the solid curve at zero level shows the residuals. The four dashed curves thus appear to reproduce the redshift distribution reasonably well. The portion of the peak below (on the blue side of) each predicted intrinsic redshift value is interpreted as evidence for the presence of ejection velocities in the measured redshifts.
The three peaks in Fig 1 also all show clear evidence for the predicted asymmetry, with the red wing of each peak extended further out than the blue wing. The dotted curves in the red wings of the two central profiles indicate the approximate extent of the ejection component in each red wing. The excess component, or \em asymmetry, \em in each red wing is assumed to be due to the cosmological component, and its extent on the high-redshift side of the peaks is a measure of the size of the Hubble flow components present in the redshifts of these sources.

The above approach is only a qualitative one, however, and the high source density in this sample may allow a more rigorous analysis. At the suggestion of the referee, a further, more quantitative analysis has been carried out on this sample using a program that fits multiple peaks in a data distribution using Chi-squared fitting. In this analysis it was assumed because of the visible asymmetry in the peaks that they were each composed of a Gaussian peak (P$_{E}$) near zero cosmological redshift, broadened by ejection velocities, plus other sources containing both an ejection component and an excess redshifted component due to the Hubble flow. The width of the P$_{E}$ peaks can then be used to estimate the value of the ejection velocities. For the purpose of this analysis the sources containing the excess Hubble flow component were also assumed to be fitted by a Gaussian (P$_{H}$), which was offset from P$_{E}$ on the high-redshift side. Although only approximate, the amount of the offset found, P$_{H}$ - P$_{E}$, can then be used as a measure of the extent of the mean Hubble flow component. By having the program fit all three Gaussian parameters for each peak, \em an independent estimate of the intrinsic redshift constant z$_{f}$ can be obtained from the redshift of each P$_{E}$ peak. \em

\begin{deluxetable}{cccccc}
\tabletypesize{\scriptsize}
\tablecaption{Gaussian Parameters from Peak Fitting in the High-Redshift Sample. \label{tbl-2}}
\tablewidth{0pt}
\tablehead{
\colhead{component} & \colhead{$N$} & \colhead{Peak Amp} & \colhead{Redshift(z)} & \colhead{HWHM}
&  \colhead{(z${f}$ = z+0.031)/$N$}
}
\startdata

1(P$_{E}$) & 4 & 501.73 & 2.460 & 0.138 & 0.623 \\
2(P$_{H}$) & 4 & 135.44 & 2.707 & 0.073 & ---  \\
3(P$_{E}$) & 5 & 320.65 & 2.985 & 0.156  & 0.603 \\
4(P$_{H}$) & 5 & 270.96 & 3.252 & 0.157 & --- \\
5(P$_{E}$) & 6 & 200.52 & 3.701 & 0.173  & 0.622  \\
6(P$_{H}$) & 6 & 68.09 & 3.993 & 0.178  &  ---  \\
7(P$_{E}$) & 7 & 76.93 & 4.325 & 0.166  & 0.622 \\  
8(P$_{H}$) & 7 & 23.56 & 4.683 & 0.157  &  --- \\
\enddata 
\end{deluxetable}


\begin{figure}
\hspace{-1.0cm}
\vspace{-1.5cm}
\plotone{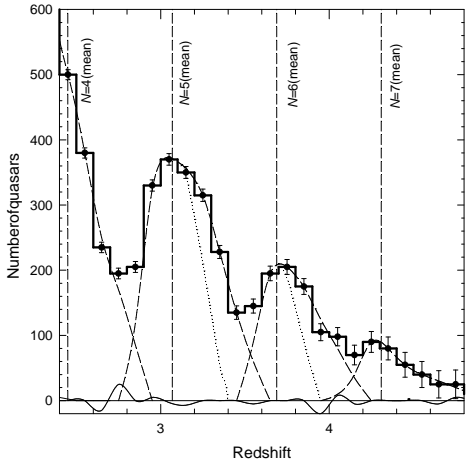}
\caption{\scriptsize{The bold histogram gives the distribution of approximately 5000 SDSS quasars between z = 2.4 and z = 4.8. Vertical dashed lines indicate the positions of predicted preferred redshifts. Dashed curves indicate the four peak profiles that reproduce the histogram. Residuals are given by the solid curve at zero level. The dotted curves separate the ejection components from the Hubble flow components in the red wing of the two central peaks. \label{fig1}}}
\end{figure}



\begin{figure}
\hspace{-1.0cm}
\vspace{-1.0cm}
\epsscale{1.0}
\plotone{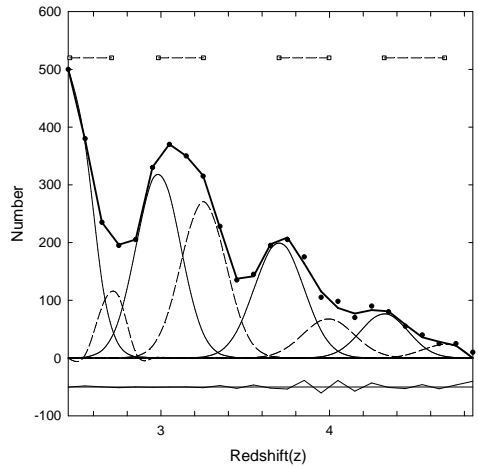}
\caption{\scriptsize{Chi-squared fit to peaks in the high-redshift sample (see text).  \label{fig2}}}
\end{figure}

The parameters found in the peak-fitting analysis are listed in Table 2 and the Gaussians are plotted in Fig 2 where the solid and dashed curves represent the P$_{E}$ and P$_{H}$ peaks respectively. Here the data are plotted as filled circles and the sum of the Gaussians is given by the bold plot. The residuals are shown in the redshift axis offset at -50. The RMS value obtained for the Chi-squared fit residuals was 5.9. In Table 2 the peak halfwidths at half maximum (HWHM) are given in column 5, and the 8 Gaussians are identified by the redshifts of the peaks which are plotted as open squares near the top of Fig 2. Peaks 1, 3, 5, and 7 represent the intrinsic redshift peaks, P$_{E}$, broadened only by their ejection velocity components. Peaks 2, 4, 6, and 8 represent the peaks, P$_{H}$, containing an additional Hubble flow component. A measure of the extent of this component is given by the length of the horizontal dashed lines in Fig 2 (the redshift difference between the relevant P$_{H}$ and P$_{E}$ peaks).

There is no reason to suspect that the mean value of the Hubble flow components should vary significantly from peak to peak. For the range of redshifts involved here (from z = 2.4 to z = 4.8), and the relation (1+z$_{c}$) = (1+z)/(1+z$_{i}$), where z is the measured redshift and z$_{c}$ and z$_{i}$ are the cosmological and intrinsic redshifts respectively, a significant increase in the separation between the P$_{E}$ and P$_{H}$ peaks is predicted with increasing redshift even if the mean Hubble flow component (z$_{c}$) remains constant from peak to peak. This increase is visible in the horizontal dashed lines in Fig 2, and in Fig 3, where the relevant P$_{H}$ - P$_{E}$ separations have been plotted as filled circles. The solid line in Fig 3 is a linear fit to these data. The dashed line shows how the separations are expected to increase for a constant z$_{c}$ = 0.066 and shows excellent agreement with the data. This mean Hubble flow component corresponds to a distance of 340 Mpc. All distances assume H$_{\rm o}$ = 58.

The z$_{f}$-values obtained from the P$_{E}$ peaks are listed in column 6 of Table 2. Since the centers of the peaks are assumed to represent the mean of the two intrinsic redshift components for each peak, the intrinsic redshift constant z$_{f}$ is obtained by adding z = 0.031 (half the intrinsic component spacing) to the redshift obtained for each P$_{E}$ peak and dividing the result by the relevant $N$ value. The mean value obtained is z$_{f}$ = 0.6175.
This value is in excellent agreement with the z$_{f}$-value found in previous work. Although the intrinsic redshift constant z$_{f}$ is almost identical to the so-called Golden Ratio (0.618), it is assumed here that this is purely coincidental. 

These results then suggest that the mean distance of non-lensed QSOs detected in the SDSS is $\sim340$ Mpc.

Similarly, from column 5 of Table 2, the half-widths of the P$_{E}$ peaks increase with z. The first three of these, where the source density is high, have been plotted in Fig 4. Here the solid line is a linear fit to the data and the dashed line corresponds to the increase expected for a constant z$_{\rm e}$ = 0.037 in the source frame. This corresponds to a mean ejection velocity of 11,000 km s$^{-1}$, which agrees well with values found in the NGC 1068 study (up to 25,000 km s$^{-1}$ \citep{bel02c}.   

The peak-fitting analysis was repeated by fixing the positions of the  P$_{E}$ peaks at the values obtained assuming z$_{f}$ = 0.62. However, the fit was poorer, resulting in a residual RMS value of 13.1 compared to 5.9 obtained above. Also, the peak spacing and width relations visible in Figs 3 and 4 were poorer. 

\begin{figure}
\hspace{-1.0cm}
\vspace{-1.0cm}
\epsscale{1.0}
\plotone{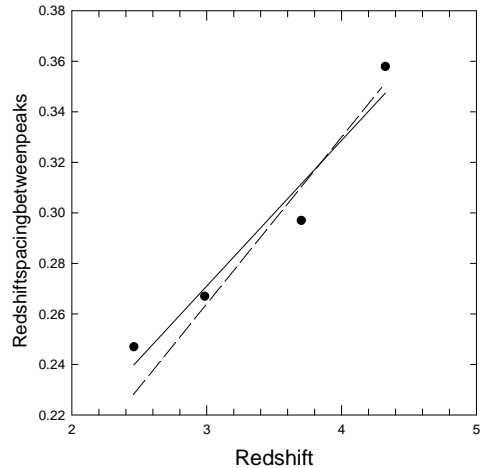}
\caption{\scriptsize{Difference between the P$_{E}$ and P$_{H}$ peaks as described in the text. The solid line is a linear fit to the data while the dashed line corresponds to sources with a constant mean cosmological distance of z$_{c}$ = 0.066.  \label{fig3}}}
\end{figure}

\begin{figure}
\hspace{-1.0cm}
\vspace{-1.0cm}
\epsscale{1.0}
\plotone{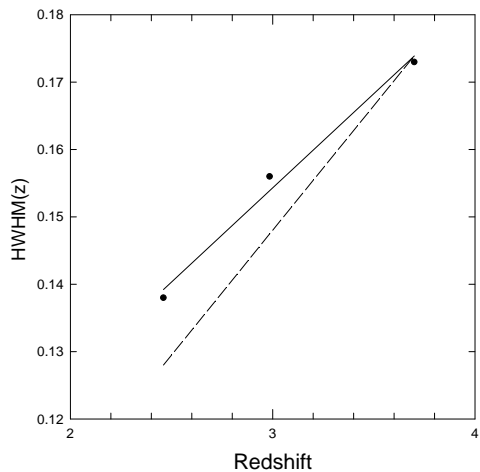}
\caption{\scriptsize{HWHM versus Redshift for the strongest three peaks in the high-redshift distribution. The dashed line corresponds to the increase expected for a constant ejection velocity of 11,000 km s$^{-1}$.   \label{fig4}}}
\end{figure}


\subsection{The Low-Redshift Sample}

The bold histogram in Fig 5 represents the redshift distribution for all objects with $0.02 < z < 0.2$ found in early surveys. They have been taken from Table 1 of \citet[see also Burbidge and Hewitt 1990]{hew93}. Prominent peaks at z = 0.031 and 0.062 are visible. Other weaker peaks at z = 0.12 and 0.18 are also present. They represent the n = 2 and n = 3 peaks reported by \citet{bur68,bur90} that are part of a series of peaks harmonically related to z = 0.06, between z = 0.06 and z = 0.6. They also represent the $N$ = 1 group of preferred values defined by eqn 1. The strong peak at z = 0.062 was pointed out by \citet{bur90}. Its position agrees with the only preferred value predicted by the log(1+z) relation \citep{kar71,kar77} to fall in this redshift range.


\begin{figure}
\hspace{-1.0cm}
\vspace{0.0cm}
\plotone{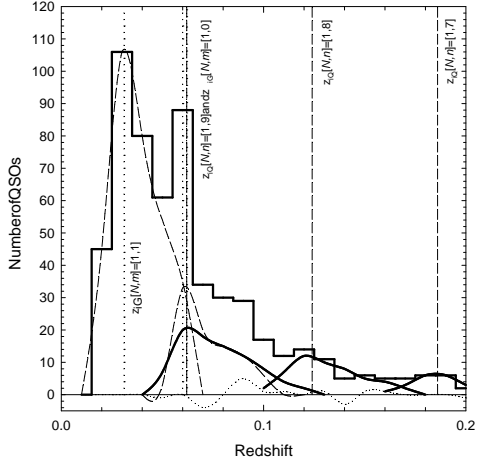}
\vspace{-2.0cm}
\caption{\scriptsize{Bold histogram shows the distribution of redshifts between z = 0.02 and 0.2 for quasars and quasar-like objects from \citet{bur90}. The two dashed curves correspond to galaxy distributions at the two galaxy preferred values above z = 0.02 that fall in this window. The three bold, solid curves correspond to peaks at the quasar preferred values. The intrinsic redshift values present in this redshift range are indicated by the vertical dashed (quasar) and dotted (galaxy) lines. The residuals are shown by the dotted curve at the zero level (see text). \label{fig5}}}
\end{figure}



\begin{figure}
\hspace{-1.0cm}
\vspace{-1.0cm}
\epsscale{1.0}
\plotone{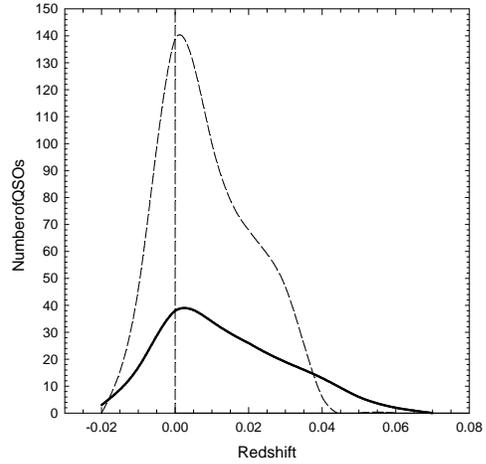}
\caption{\scriptsize{Solid and dashed curves show the redshift distributions in Fig 5 for the low-redshift sample of quasars and quasar-like galaxies respectively, after removal of the intrinsic components.  \label{fig6}}}
\end{figure}

\begin{deluxetable}{cccccc}
\tabletypesize{\scriptsize}
\tablecaption{Peak Fitting Parameters for Quasar-like Galaxies. \label{tbl-3}}
\tablewidth{0pt}
\tablehead{
\colhead{component} & \colhead{Peak Amp} & \colhead{Redshift(z)} & \colhead{Distance(Mpc)} & \colhead{HWHM} & \colhead{Eject. Vel.(km s$^{-1}$)}
}
\startdata
Gal(P$_{E}$) & 150.1 & 0.0019 & 9.8 & 0.0085 & 2550 \\
Gal(P$_{H}$) & 68.04 & 0.023 & 120 & 0.0091 & 2730 \\
Quasar(P$_{E}$) & 31.14 & 0.0015 & 7.8 & 0.011 & 3300  \\
Quasar(P$_{H}$) & 20.8 & 0.024 & 124 & 0.019 & 5700 \\
\enddata 
\end{deluxetable}

In Fig 5 the vertical dashed lines represent the location of the preferred values predicted by eqn 1, and the dotted lines represent the preferred values predicted by the galaxy equation (eqn B1 of \citet{bel03b}), for the most common, $N$ = 1, case. Each line is identified. They are all located at peaks in the redshift distribution. The bold, solid curves represent three similar-shaped profiles, obtained qualitatively as in Fig 1, that correspond to the three quasar intrinsic redshift peaks visible in the histogram. The two dashed curves are similar shaped curves located at the galaxy peaks at z = 0.031 and 0.062. The dotted curve at the zero level indicates the residual, or difference, between the histogram and the sum of the three solid (quasar) and two dashed (galaxy) curves. Because of the enhanced region near z = 0.09 on the histogram, the profile shown (with an extended red wing) is the only one found that would keep the residuals low. It is not known if there are duplicate preferred values at z = 0.062 (one for quasars and one for galaxies) or just one. However, both cases will lead to the same residuals if the two z = 0.062 peaks shown are combined into one.



\section{The Radial Extent of Detected Quasars and Quasar-Like Galaxies}

In Fig 6 the redshift distribution in Fig 5 has been re-plotted after removal of the intrinsic components. The dashed and solid curves are for galaxies and quasars respectively. Note that, although some of the objects with redshifts above z = 0.062 may be quasar-like galaxies, they are referred to here as quasars since their preferred redshifts are predicted by eqn 1. In the DIR model objects are expected to look similar if they have similar intrinsic redshift components, since the intrinsic redshift is interpreted as a direct indication of their age.

The asymmetry in the curves in Fig 6 is clear, with the red wing extended much further out than the blue wing. In these curves, as in Fig 1, the blue wing is assumed to be an indication of the ejection velocities. Since these sources are assumed to be much older than those making up the high-redshift sample, their present velocities (peculiar velocities) are expected to be much smaller because of adiabatic expansion. In addition to this component, the red wing is further broadened by an additional Doppler component due to the Hubble flow. Because of the low number of sources involved no attempt was made to fit Gaussians to the many peaks in Fig 5, however, double Gaussians were fitted to each of the two curves in Fig 6 and the results are given in Table 3. The fits gave residual RMS values of 0.5 and 3.3 for the quasar and quasar-like galaxy data respectively. Note that the peaks near zero redshift found for both the quasars and quasar-like galaxies are slightly offset at 7.8 and 9.8 Mpc respectively. The radial extent of both groups is similar, as expected, but is now only 122 Mpc. This is much smaller than the 340 Mpc extent found for the high-redshift sample, but it is completely expected because of the more sensitive detection limit of the SDSS.

The mean ejection velocities are listed in col 6 of Table 3 and, as expected, are significantly smaller than the 11,000 km s$^{-1}$ value found for the high redshift (large intrinsic redshift) sample. As discussed above, this decrease in peculiar velocity with decreasing intrinsic redshift (increasing age) is anticipated in the DIR model, assuming adiabatic expansion, since the high and low-redshift objects have ages that range from 10$^{7}$ to 10$^{9}$ yrs respectively.

The range of distances found here is in reasonable agreement with that reported recently by \citet{bur04}.

\section{Discussion}

Much evidence has been presented here and elsewhere that supports the DIR model and the ejection of QSOs from the nuclei of active galaxies.  However, there are also some well known arguments against this model that are raised by those who support the cosmological redshift model, and these need to be at least briefly addressed. 


When high-redshift quasars are gravitationally lensed by intervening galaxies it is argued that this indicates that quasar redshifts must be cosmological. However, this argument only rules out models in which all quasars are located nearby. In the DIR model, although intrinsically somewhat fainter, quasars are uniformly distributed throughout the Universe and there is every reason to suspect that some of the distant ones will be sufficiently magnified by the lensing process to be detectable. Of the 82 lensing cases listed in \citet{koc04}, redshifts are known for 43 of the intervening lenses. The mean lens redshift is z = 0.55. Furthermore, in most cases there is insufficient evidence known about the lensing object to rule out quasar-like galaxies as the lens galaxy, which can also have a small intrinsic component. There is therefore no reason to believe that lensed quasars behind these objects cannot be detected when magnified. Also, recent searches for lensing cases amongst very high redshift quasars have not produced the expected number \citep{ric04}, even though one expects that an appreciable fraction of these objects will be gravitationally lensed if they are at cosmological distances. The sample may still be too small, however, for meaningful conclusions to be drawn. 


The damped Ly${\alpha}$ systems in quasars contain a small amount of dust and heavy elements that may be the progenitors of galactic disks. This picture is not in conflict with the DIR model. Narrow Ly-$\alpha$ absorption lines are observed on the short wavelength side of broad Ly-$\alpha$ emission lines in high redshift quasars. This so-called Ly$\alpha$ forest has conventionally been interpreted as being due to absorption in galaxies that lie along the line-of-sight to the quasar. However, in the DIR model, \em depending on how the intrinsic redshifts are produced, \em matter near the QSO might also produce Ly$\alpha$-like absorption features over a large range of redshifts even though this matter is all located relatively close to the QSO, if it too has an intrinsic component, is being driven out of the QSO toward the observer, or both. The Ly$\alpha$ forest is seen \em predominantly \em in quasars, and quasar-like galaxies. Since the redshifts of normal galaxies are, without question, almost entirely cosmological, detection of the same Ly$\alpha$ forest in these galaxies would then seem to be an excellent way of confirming that quasars are also at cosmological distances. Unfortunately this has not happened convincingly and the interpretation of the Ly$\alpha$ forest is still being questioned \citep{han03}. In fact, there appears to be good evidence to show that these features are not seen in galaxies \citep{col02}, which is a strong argument in favor of the DIR model. There also appears to be evidence that the element abundances in the damped Ly$\alpha$ systems are surprisingly much more uniform than would be expected for intervening galaxies \citep{pro02}. The lack of change in average metallicity with redshift is also difficult to understand if quasars are at cosmological distances, because evolution since the Big Bang is expected to build up more metals with each new generation of stars. This lack of change in metallicity is exactly what is expected in the DIR model. This evidence therefore may be more consistent with the Ly$\alpha$ systems \em not \em being produced by ordinary galaxies lying along the line-of-sight to these objects.

Also, there is no question that the power spectrum of the redshifts of galaxies and quasars is potentially an important means of probing preferred redshifts.
An analysis of the QSO power spectrum of 22,652 QSOs in the 2dF QSO Redshift Survey catalogue has been presented by \citet{out03} who found,
assuming cosmological redshifts, that the amplitude of clustering of QSOs at a mean redshift of z = 1.4 is similar to that of present day galaxies. A comparison of the power spectra of QSOs at low and high redshift also showed little evidence for any evolution in the amplitude.
The spatial plots of galaxies, such as those in Figs 4 and 5 of \citet{teg04} show clearly the density structure of galaxies in the local universe. Similar plots are available for the 2dF galaxy data. Unfortunately similar plots for quasars are hard to find. It would seem that this type of structure should also be visible in quasar plots if the redshifts are truly cosmological. Presumably \citet{out03} do not include such a plot because the source density per unit volume in space is too low to show structure (if the 22,600 quasars extend to redshifts near z = 3 and the 205,000 galaxies plotted by \citet{teg04} extend only to z = 0.2, the galaxy/quasar number density ratio per unit volume between these two samples is huge ($\sim8000/1$). Such a plot \em is \em presented for the SDSS Early Data Release for quasars \citep{sto02}. There is little evidence in the QSO plot for the type of structure seen in the galaxy plots, however, this may simply mean that the QSO source density is still too low as suggested above. Although it is true that the QSO power spectrum results of \citet{out03} agreed with galaxy results it is not clear that the discrete features in the redshift distribution that we are discussing here could not have influenced the result. Furthermore, it is difficult to prove that the DIR model would not also give acceptable results since the redshift distribution would be completely new and only by repeating the analysis using the new distance redshifts of the DIR model could this be determined. Unfortunately these redshifts remain unknown. It should also be kept in mind that the results obtained in this paper were obtained using only raw data (redshifts), and the analysis, unlike the power spectrum analysis, does not depend on involved calculations and assumptions. Although power spectrum analysis is potentially a powerful technique for examining the spatial structure of galaxies in the Universe, it may still be too early for quasar results to be meaningful. 

\section{Conclusion}

Using two quasar samples, one with high redshifts and one with low, it has been shown that not only do all peaks in these redshift distributions occur at previously predicted preferred values, they also all show evidence for an extended red wing suggesting the presence of small cosmological components in an otherwise intrinsic redshift distribution. This result is shown to be consistent with the decreasing intrinsic redshift model described in several previous papers, where QSOs with a high intrinsic redshift component are assumed to be ejected from the nuclei of active galaxies. For the low and high redshift samples studied here the mean cosmological components are found to be z$_{c}$ $\sim0.024$ and $\sim0.066$ respectively, with the difference explained by the different detection limits of the two samples. A continuous reduction in the size of the ejections velocities (peculiar velocites) of quasars is seen as the intrinsic redshift component decreases (i.e. as they age). These results are further evidence in favor of models proposing that quasars are ejected from active galaxies. 

I thank an anonymous referee for suggestions that have significantly improved this paper.



\end{document}